%% file: bscorr.tex
\documentstyle[11pt]{article}
\input defn

\newcommand{\INIT}{{\em INIT}}
\newcommand{\INITp}{\mbox{{\em INIT}$\,'$}}
\newcommand{\Rrep}{{\bf R}}
\newcommand{\Pg}{{\sf Pg}}
\newcommand{\Pgkb}{{\sf Pg}_{{\it kb}}}

\newcommand{\DIFF}{{\sf DIFFUSE}}
\newcommand{\citeyear}{\cite}
\newcommand{\noop}{\mbox{{\sf no--op}}}

\begin{document}
\begin{titlepage}
\title{A Note on Knowledge-Based Programs and Specifications}
\author{
Joseph Y.\ Halpern%
\thanks{Much of this work was carried out while the author was at the
IBM Almaden Research Center.  IBM's support is gratefully acknowledged.
The work was also supported in part by NSF under
grant IRI-96-25901, and by the Air Force Office of
Scientific Research under contract
 F49620-91-C-0080 and grant F49620-96-1-0323.}\\
Computer Science Dept.\\
Cornell University \\
   Ithaca, NY 14853\\
   halpern@cs.cornell.edu\\
   http://www.cs.cornell.edu/home/halpern}

\date{ }
\maketitle
\thispagestyle{empty}

\begin{abstract}
Knowledge-based program are programs with explicit tests for knowledge.
They have been used successfully in a number of applications.  Sanders
has pointed out what seem to be a counterintuitive property of
knowledge-based programs.  Roughly speaking, they do not satisfy a
certain monotonicity property, while standard programs (ones without
tests for knowledge) do.  It is shown that there are two ways of
defining the monotonicity property, which agree for standard programs.
Knowledge-based programs satisfy the first, but do not satisfy the
second.  It is further argued by example that the fact that they do not
satisfy the second is actually a feature, not a problem.  Moreover, once
we allow the more general class of {\em knowledge-based
specifications}, standard programs do not satisfy the monotonicity
property either.
\end{abstract}

\end{titlepage}
\section{Introduction}
\nocite{FHMV}
Consider a simple program such as
\begin{figure}[htb]
$$\begin{array}{l}
{\bf do\ forever}\\
\ \ \ \ \ \mbox{{\bf if} $x=0$ {\bf then} $y := y+1$ {\bf end}}\\
{\bf end}.
\end{array}$$
\caption{The program $\Pg_1$}\label{fig1}
\end{figure}
This program, denoted $\Pg_1$ for future reference,
describes an action that a process (or agent---I use the two words
interchangeably here)
should take, namely, setting $y$ to $y+1$, under certain conditions,
namely, if $x=0$.  One way to way to provide formal semantics for such
a program is to assume that each agent is in some {\em
local state\/}, which, among other things, describes the value of the
variables of interest.  For this simple program, we need to assume that
the local state contains enough information to determine the truth of
the test $x=0$.
We can then associate with the program a {\em
protocol}, that is, a function describing what action the agent should
in each local state.  Note that a program is a {\em syntactic\/} object,
given by some program text, while a protocol is a function, a {\em
semantic\/} object.

{\em Knowledge-based programs},
introduced in \cite{FHMV,FHMV94}
(based on the {\em knowledge-based protocols\/} of \cite{HF87})
are intended to provide a
high-level framework for the design
and specification of protocols.  The idea is that, in knowledge-based
programs, there are explicit tests for knowledge.  Thus, a
knowledge-based program might have the form
\begin{figure}
$$\begin{array}{l}
{\bf do\ forever}\\
\ \ \ \ \ \mbox{{\bf if} $K(x=0)$ {\bf then} $y := y+1$ {\bf end}}\\
{\bf end},
\end{array}$$
\caption{The program $\Pg_2$}\label{fig2}
\end{figure}
where $K(x=0)$ should be read as ``you know $x=0$''.
We can informally view this knowledge-based program, denoted $\Pg_2$,
as saying ``if you know that $x=0$, then set $y$ to $y+1$''.
Roughly speaking, an agent knows $\phi$ if, in all situations consistent
with the agent's information, $\phi$ is true.

Knowledge-based programs
are an attempt to capture the intuition that what
an agent does depends on what it knows.  They have already
met with some degree of success, having been used
in papers such as \cite{DM,Had,HMW,HZ,Maz,Maz90,MT,NT} both to
help in the design of new protocols and to clarify the understanding of
existing protocols.
However, Sanders \citeyear{Sanders} has pointed out what seems to be a
counterintuitive property of knowledge-based programs.  Roughly
speaking, she claims that knowledge-based programs do not satisfy a
certain monotonicity property:
a knowledge-based program can satisfy a specification under a given
initial condition, but fail to satisfy it if we strengthen the initial
condition.  On the other hand, standard programs (ones without tests for
knowledge) do satisfy the monotonicity property.

In this paper, I consider Sanders' claim more carefully.  I show that
it depends critically on what it means for a program to satisfy
a specification.  There are two possible definitions, which agree for
standard programs.  If we use the one closest in spirit to the ideas
presented in \cite{HF87}, the claim is false, although it is true for
the definition used by Sanders.  But, even in the case of Sanders'
definition, rather than being a defect of
knowledge-base programs, this lack of monotonicity is actually a
feature. In general, we
do not want monotonicity.  Moreover, once we allow a more
general class of {\em knowledge-based specifications}, then standard
programs do not satisfy the monotonicity property either.

The rest of this paper is organized as follows:  In the next section,
there is an informal review of the semantics of standard
and knowledge-based
programs.  In Section~\ref{specs}, I discuss standard and
knowledge-based specifications.
In Section~\ref{monotonicity}, I consider the monotonicity property
described by Sanders, and show in what sense it is and is not satisfied
by knowledge-based programs.  I give some examples in
Section~\ref{examples} showing why monotonicity is not always desirable.
I conclude in Section~\ref{conclusion} with some discussion of
knowledge-based programs and specifications.

\section{Standard and knowledge-based programs: an informal
review}
Formal semantics for standard and knowledge-based programs are provided
in
\cite{FHMV,FHMV94}.
To keep the discussion in this paper at an informal
level, I simplify things somewhat here, and review
what I hope will be just enough of the details
so that, together with the examples given here, the reader will be able
to follow the main points; the interested
reader should refer to \cite{FHMV,FHMV94} for
further discussion and all the formal details.

Informally, we view a distributed system as consisting of a number
of interacting agents.  We assume that, at any given point in time,
each agent in the system is in some {\em local state}.
A {\em global state\/} is just a
tuple consisting of each agent's local state,
together with the state of the {\em environment}, where the
environment consists of everything that is relevant to the system that
is not contained in the state of the processes.  The agents' local
states typically change over time, as a result of actions that they
perform.  A {\em run\/} is a function from time to global states.
Intuitively, a run is a complete description of what happens over time
in one possible execution of the system.  A {\em point\/} is a pair
$(r,m)$ consisting of a run $r$ and a time $m$.  At a point $(r,m)$, the
system is in some global state $r(m)$.  For simplicity, time
here is taken to range over the natural numbers (so that time is viewed
as discrete, rather than continuous).   A {\em system\/} $\R$ is
a set of runs; intuitively, these runs describe all the
possible executions of the system.  For example, in a poker game, the
runs could describe all the possible deals and bidding sequences.

Of major interest in this paper are the systems that we can associate
with a program.  To do this, we must first associate a system with a
{\em joint protocol}.   As was said in the introduction, a protocol is a
function from local states to actions.   (This function may be
nondeterministic, so that in a given local state, there is a set of
actions that may be performed.)  A joint protocol is just a set of
protocols, one for each process.

While the joint protocol describes what each process does, it does not
give us enough information to generate a system.  It does not tell us
what the legal behaviors of the environment are, the effects of the
actions, or the initial conditions.  We specify these in the
{\em context}.  Formally, a context $\gamma$ is a tuple
$(P_e,\Gz,\tau,\Psi)$, where $P_e$ is a protocol for the environment,
$\Gz$ is a set of initial global states,
$\tau$ is a {\em transition function}, and $\Psi$ is a set of
{\em admissible\/} runs.
The environment is viewed as running a protocol just like the agents;
its protocol is used to capture features of the setting like
``all messages are delivered within 5 rounds'' or ``messages may be
lost''.  Given a joint protocol $P = (P_1, \ldots, P_n)$ for the agents,
an environment protocol $P_e$, and a global state $(s_e, s_1, \ldots,
s_n)$, there is a set of possible {\em joint actions\/} $({\sf a}_e,
{\sf a}_1, \ldots, {\sf a}_n)$ that can be performed in this global
state according to the protocols of the agents and the environment.  (It
is a set since the protocols may be nondeterministic.)
The transition function $\tau$
describes how these joint actions
change the global state by associating with each joint
action a {\em global state
transformer}, that is, a mapping from global states to global states.
The set $\Psi$ of admissible runs is used to
characterize notions like fairness.
For the simple programs considered in this paper, the transition
function will be almost immediate from the description of the global
states and $\Psi$ will typically consist of all runs (so that it
effectively plays no interesting role).   What will
change as we vary the context is the set of possible initial global
states.

A run $r$ is {\em consistent with joint protocol $P$ in context
$\gamma$\/} if (1) $r(0)$, the initial global state of $r$, is one of
the initial global states in $\Gz$, (2) for all $m$, the transition
from global state $r(m)$ to $r(m+1)$ is the result of applying $\tau$ to
a joint action that can be performed by $(P_e,P)$
in the global state $r(m)$, and (3) $r \in \Psi$.
A system $\R$ {\em represents\/} a joint protocol $P$ in context
$\gamma$ if it consists of all runs consistent with $P$ in
$\gamma$.

Assuming that each test in a standard program run by process $i$ can be
evaluated in each local state, we can derive a
protocol from the program in an obvious way: to find out what process $i$
does in a local state $\ell$,
we evaluate the
tests in $\Pg$ in $\ell$ and perform the appropriate action.%
\footnote{Strictly speaking, to evaluate the tests, we need an {\em
interpretation\/} that assigns truth values to formulas in each global
state.  For the programs considered here, the
appropriate interpretation will be immediate from the description of the
system, so I ignore interpretations here for ease of exposition.}
A run is {\em consistent with $\Pg$ in context $\gamma$\/} if it is
consistent with the protocol
derived from $\Pg$.  Similarly,
A system {\em represents $\Pg$ in
context $\gamma$\/} if it represents the protocol derived from $\Pg$.  We use $
\Rrep(\Pg,\gamma)$ to denote the system representing $\Pg$ in context
$\gamma$.

\xam\label{pgxam} Consider
the simple standard program $\Pg_1$ in Figure~\ref{fig1}
and suppose there is only one agent in the
system.  Further suppose the agent's local state is a pair
of natural numbers $(a,b)$, where $a$ is the current value of
variable $x$ and $b$ is the current value of $y$.
The protocol derived from $\Pg_1$ increments the value of $b$ by 1
precisely if $a=0$.
In this simple case, we can ignore the environment state,
and just identify the global state of the system with the agent's local
state.  Suppose we consider the context $\gamma$ where the initial
states consist
of all possible local states of the form $(a,0)$ for $a \ge 0$
and the transition function is such
that the action $y := y+1$ transforms $(a,b)$ to $(a,b+1)$.
We ignore the environment protocol (or, equivalently, assume that $P_e$
performs the action $\noop$ at each step) and assume $\Psi$ consist of
all runs.
 A run $r$ is then consistent with $\Pg_1$ in context $\gamma$
if either (1) $r(0)$ is of the form $(0,b)$ and $r(m)$ is of the form
$(0,b+m)$ for all $m \ge 1$, or (2) $r(m)$ is of the form $(a,b)$ for
all $m$ and $a > 0$.  That is, either the $x$ component is originally
0, in which case the $y$ component is continually increased by 1, or
else nothing happens. \exam

Now we turn to knowledge-based programs.
Here the situation
is somewhat more complicated.
In a given context,
a process can determine the truth of a
test such as ``$x=0$'' by simply looking at its local state.
However, in a knowledge-based program, there are tests for knowledge.
According to the definition of knowledge in systems,
an agent $i$ knows a fact
$\phi$ at a given point $(r,m)$ in system $\R$ if $\phi$ is
true at all points in $\R$ in which $i$ has the same local state as it
does at $(r,m)$.  Thus, $i$ knows $\phi$ at the point $(r,m)$ if $\phi$
holds at all points consistent with $i$'s information at $(r,m)$.
The truth of a test for knowledge cannot in general
be determined simply by
looking at the local state in isolation.  We need to look at the whole
system.  As a consequence,
given a run, we cannot in general determine if it is consistent with a
knowledge-based program in a given context.  This is because we cannot
tell how the tests for knowledge turn out without
being given the other possible runs of the system; what a process knows
at one point will depend in general on what other points are possible.
This stands in sharp contrast to the situation for standard programs.

This means it no longer makes sense to talk about a run being consistent
with a knowledge-based program in a given context.
However, notice that, given a system $\R$, we
can derive a protocol from a knowledge-based program $\Pgkb$ for
process $i$ by using
$\R$ to evaluate the knowledge tests in $\Pgkb$.  That is, a
test such as $K \phi$ holds in a local state $l$ if $\phi$ holds at all
points in $\R$ where process $i$ has local state $l$.  In general,
different protocols can be derived from a given knowledge-based program,
depending on what system we use to evaluate the tests.  Let $\Pgkb^\R$
denote the protocol derived from $\Pgkb$ given system $\R$.

We say that a system $\R$ {\em represents\/} a knowledge-based program
$\Pgkb$ in context $\gamma$ if $\R$ represents the protocol $\Pgkb^\R$.
That is, $\R$ represents $\Pgkb$ if $\R =
\Rrep(\Pgkb^\R,\gamma)$.  Thus, a system represents $\Pgkb$ if it
satisfies a certain fixed-point equation.

This definition
is somewhat subtle, and determining the system representing a given
knowledge-based program may be nontrivial.
Indeed, as shown in \cite{FHMV,FHMV94},
in general, there may be no systems representing a knowledge-based
program $\Pgkb$ in a given context, only one, or more than one, since
the fixed-point equation may have no solutions, one solution, or many
solutions. Moreover,
computing the solutions may be a difficult task, even if we have only
finitely many possible global states.  There are conditions
sufficient to guarantee that there is exactly one
system representing $\Pgkb$, and these conditions are satisfied by many
knowledge-based programs of interest, and, in particular, by the
programs discussed in this paper.  If $\Pgkb$ has a
unique system representing it in context $\gamma$, then we again denote
this system $\Rrep(\Pgkb,\gamma)$.

\xam\label{pgxam2}
The knowledge-based program $\Pg_2$ in Figure~\ref{fig2},
with the test $K(x=0)$, is particularly simple to analyze.  If we
consider the context $\gamma$ discussed in Example~\ref{pgxam}, then
whether or not $x=0$ holds is determined
by the process' local state.  Thus, in context $\gamma$, $x=0$ holds
iff $K(x=0)$
holds, and the knowledge-based program reduces to the standard program.

On the other hand, consider the context $\gamma'$ where the agent's
local state just
consists just of the value of $y$, while the
value of $x$ is part of the environment state.  Again, we can identify
the global state with a pair $(a,b)$, where $a$ is the current value of
$x$ and $b$ is the current value of $y$, but now $a$ represents the
environment's state, while $b$ represents the agent's state.  We can
again assume the environment performs the $\noop$ action at each step,
$\Psi$ consists of all runs, the transition function is as in
Example~\ref{pgxam}, and the initial states are all possible global
states of the form $(a,0)$.   In this context, there is a also unique
system representing
$\Pg_2$:  The agent never knows whether $x=0$, so there is a unique run
corresponding to each initial state $(a,0)$, in which the global state
is $(a,0)$ throughout the run.

Finally, let $\gamma''$ be identical to $\gamma'$ except that the only
initial state is $(0,0)$.  Again, there will be a unique system
representing $\Pg_2$ in $\gamma''$, but it is quite different from
$\Rrep(\Pg_2,\gamma')$.   In $\Rrep(\Pg_2,\gamma'')$, the agent knows
that $x=0$ at all times.  There is only one run, where the value of $y$
is augmented at every step. \exam

This discussion suggests that a knowledge-based program can be viewed
as specifying a set of systems, the ones that satisfy a certain
fixed-point property, while a standard program can be viewed as
specifying a set of runs, the ones consistent with the program.

\section{Standard and knowledge-based specifications}\label{specs}
Typically, we think of a protocol being designed to satisfy
a {\em specification}, or set of properties.
Although a specification is often written in some specification
language (such as temporal logic), many
specifications can usefully be viewed
as predicates on runs.  This means that we can
associate a set
of runs with a specification;
namely, all the runs that satisfy the required properties.
Thus, a specification such as ``all processes eventually decide
on the same value'' would be associated with the set of runs
in which the processes do all decide the same value.%
\footnote{Of course, there are useful specifications
that cannot be viewed as predicates on runs.  While {\em linear time\/}
temporal logic assertions are predicates on runs, {\em branching time\/}
temporal logic assertions are best viewed as predicates on trees.
(See \cite{EH2,Lam80} for a discussion of the differences between
linear time and branching time.)  For example,
Koo and Toueg's notion of {\em weak
termination\/} \cite{KooToueg} requires that at every point there is
a possible future where everyone terminates.  In the notation used in
this paper, this would mean that for every point $(r,m)$, there must
be another point $(r',m)$ such that $r$ and $r'$ are identical up to
time $m$, and at some point $(r',m')$ with $m' \ge m$, every process
terminates.  This assertion is easily expressed in branching time
logic.  Probabilistic assertions such as ``all processes terminate
with probability .99'' also cannot be viewed as predicates on
individual runs.  Other examples of specifications that cannot
be viewed as a predicate on runs are discussed later in this section.
Nevertheless, specifications that are predicates on runs are
sufficiently prevalent that it seems reasonable to give them special
attention.}

Researchers have often focused attention on two
types of specifications:
{\em safety properties}---these are invariant properties that
have the form ``a particular bad thing never happens''---and
{\em liveness properties}---these are properties that essentially
say ``a particular good thing eventually does happen'' \cite{OL}.
Thus, a run $r$ has a safety property $p$ if $p$ holds at
all points $(r,m)$, while $r$ has the liveness property $q$ if
$q$ holds at some point $(r,m)$.
Suppose
we are interested in a program that guarantees that
all the processes
eventually decide on the same value.
We model this by assuming that each process $i$ has a decision
variable $x_i$, initially undefined,
in its local state
(we can assume a special ``undefined'' value in the domain),
which is set once in the course of a run,
when the decision is made.
Given the way we have chosen to model this
problem, we would expect this program to satisfy
two safety properties: (1)
each process' decision variable
is changed at most once (so that it is never the case that it is set
more than once); and (2) if neither $x_i$ nor $x_j$ has value
``undefined'', then they are equal.  We also expect it to satisfy
one liveness property:
each decision variable is eventually set.

We say that a standard program $\Pg$ {\em satisfies\/} a specification
$\sigma$ in a context $\gamma$ if every run consistent with $\Pg$
in $\gamma$ (that is, every run in the system representing $\Pg$ in
$\gamma$) satisfies $\sigma$.  Similarly, we can say that a
knowledge-based program $\Pgkb$ satisfies specification $\sigma$ in
context $\gamma$ if every run in every system representing $\Pgkb$
satisfies $\sigma$.

The notion of specification we have considered so far can be thought of
as being {\em run based}.  A specification $\sigma$ is a predicate on (i.e.,
set of) runs and a program satisfies $\sigma$ if each run consistent with
the program is in $\sigma$.  Although run-based specifications arise
often in practice, there are reasonable specifications that are not run
based.  There are times that it is best to think
of a specification as being, not a predicate on runs, but a predicate on
entire {\em systems}.  For example, consider a knowledge base (KB) that
responds to queries by users.  We can imagine a specification that says
``To a query of $\phi$, answer `Yes' if you know $\phi$, answer `No' if
you know $\neg \phi$, otherwise answer `I don't know'.''  This
specification is given in terms of the KB's knowledge, which depends on
the whole system and cannot be determined by considering individual runs
in isolation.  We call such a specification a {\em knowledge-based
specification}.  Typically, we think of a knowledge-based specification
being given as a formula involving operators for knowledge and time.
Formally, it is simply a predicate on (set of) systems.  (Intuitively,
it consists of all the systems where the formula is valid---i.e., true
at every point in the system.)%
\footnote{As the examples discussed in Footnote~2
show, not all predicates on systems can be expressed in terms of
formulas involving knowledge and time.
I will not attempt to characterize here the ones that can be so
expressed.  It is not
even clear that such a characterization is either feasible or useful.}

We can think of a run-based
specification $\sigma$ as a special case of a knowledge-based
specification.
It consists of all those systems all of whose runs satisfy $\sigma$.
A (standard or knowledge-based) program $\Pg$ satisfies a
knowledge-based specification $\sigma$ in context $\gamma$ if every
system representing $\Pg$ in $\gamma$ satisfies the specification.

Notice that knowledge-based specifications bear the same relationship
to (standard) specifications as knowledge-based programs bear to
standard programs.   A knowledge-based specification/program in
general defines a set of systems; a standard specification/program
defines
a set of runs (i.e., a single system).

\section{Monotonicity}\label{monotonicity}
Sanders \citeyear{Sanders} focuses on a particular monotonicity property
of specifications.  To understand this property, and Sanders' concerns,
we first need some definitions.  Given contexts $\gamma =
(P_e,\Gz,\tau,\Psi)$ and $\gamma' = (P_e',\Gz',\tau',\Psi')$,
we write $\gamma'
\sqsubseteq \gamma$ if $P_e = P_e'$, $\Gz' \subseteq \Gz$, $\tau =
\tau'$, and $\Psi' \subseteq \Psi$.  That is, in $\gamma'$ there may be
fewer initial states and fewer admissible runs, but otherwise $\gamma$
and $\gamma'$ are the same.  The following lemma is almost immediate
from the definitions.

\lem\label{subset}
If $\gamma' \sqsubseteq \gamma$, then for all protocols $P$, every
run consistent with $P$ in $\gamma'$ is also consistent with $P$ in
$\gamma$, so $\Rrep(P,\gamma') \subseteq \Rrep(P,\gamma)$.
Similarly, for every standard program $\Pg$, we have
$\Rrep(\Pg,\gamma') \subseteq \Rrep(\Pg,\gamma)$. \elem

The restriction in Lemma~\ref{subset} to {\em standard\/} programs
is necessary. It is not true for knowledge-based programs.
The set of systems consistent with
a knowledge-based program can be rather arbitrary, as
Example~\ref{pgxam2} shows.  This example also shows that safety and
liveness properties need not be preserved when we restrict the context.
The safety property ``$y$ is never equal to 1'' is satisfied
by $\Pg_2$ in context $\gamma'$ but not in context $\gamma''$.  On the
other hand, the liveness property ``$y$ is eventually equal to 1'' is
satisfied by $\Pg_2$ in context $\gamma''$ but not $\gamma'$.

Sanders suggests that this behavior is somewhat
counterintuitive.  To quote \cite{Sanders}:
\begin{quote}
[A] {\em knowledge-based protocol need not be monotonic with respect to
the initial conditions} \ldots
[In particular,] {\em safety and liveness properties of knowledge-based
protocols need not be preserved by strengthening the initial
conditions}, thus violating one of the most intuitive and fundamental
properties of standard programs [italics Sanders'].%
\footnote{In \cite{HF87}, a notion of knowledge-based {\em
protocol\/} was introduced, and Sanders is referring to that notion,
rather than the notion of knowledge-based {\em program\/} that I am
using here.  See \cite{FHMV94} for a discussion of the difference
between the two notions.  Sanders' comments apply without change to
knowledge-based programs as defined here.}
\end{quote}

It is certainly true that the system representing a
knowledge-based program in a restricted context is not necessarily a
subset of the system representing it in the original context.  However,
under what is arguably the most natural interpretation of what it means
for a program to satisfy a specification with respect to an initial
condition, a knowledge-based program
{\em is\/} monotonic with respect to initial conditions.

To understand why this should be so, we need to make precise what it
means for a (knowledge-based) program to satisfy a specification with
respect to an initial condition.  Formally, we can take an initial
condition to be a predicate on global states (so that an initial
condition corresponds to a set of global states).  An initial condition
$\INITp$ is a {\em strengthening\/} of $\INIT$ if $\INITp$ is a subset
of $\INIT$.  (In logical terms, this means that $\INITp$ can be thought
of as implying $\INIT$.)  A set $G$ of global states satisfies an
initial condition $\INIT$ if $G \subseteq \INIT$.

Suppose that we fix $P_e$, $\tau$, and $\Psi$, that is, all the
components of a context except the set of initial global states, and
consider the family $\Gamma = \Gamma(P_e,\tau,\Psi)$ of contexts of the
form
$(P_e,\Gz,\tau,\Psi)$, where the set $\Gz$ varies over all subsets of
global states.  Now it seems reasonable to say that program $\Pg$
{\em satisfies specification $\sigma$ (with respect to $\Gamma$) given
initial condition INIT\/}
if $\Pg$ satisfies $\sigma$ in every context in $\Gamma$
whose initial global states
satisfy $\INIT$. With this definition, it is clear that if $\Pg$
satisfies $\sigma$ given $\INIT$, and $\INITp$ is a strengthening of
$\INIT$, then $\Pg$ must also satisfy $\sigma$ with respect to $\INITp$,
since every context whose initial global states are in $\INITp$ also has
its initial global states in $\INIT$.

Thus, under this definition of what it means for  a program to
satisfy a specification, Sanders'
observation is incorrect.  However, Sanders used a somewhat different
definition.  Suppose that
rather than considering all contexts in $\Gamma$ whose initial
global states satisfy $\INIT$, we consider the maximal one, that is, the
one whose set of initial global states consists of all global states in
$\Sigma$ that satisfy $\INIT$.  We say that $\Pg$
{\em maximally\/} satisfies specification $\sigma$ (with respect to
$\Gamma$) given $\INIT$
if $\Pg$ satisfies $\sigma$ in the context in $\Gamma$ whose set of
initial global states consists of all global states satisfying $\INIT$.

It is almost immediate from Lemma~\ref{subset} and the definitions that
for standard programs and standard specifications, ``satisfaction with
respect to $\Gamma$'' coincides with ``maximal satisfaction with respect
to $\Gamma$''.  On the other hand, they can
be quite different for knowledge-based programs and knowledge-based
specifications, as the following examples show.

\xam\label{differ1}
For the knowledge-based program $\Pg_2$, if we take
$\Gamma$ to consist of all contexts
$(P_e,\Gz,\tau,\Psi)$, where $P_e$, $\tau$, and $\Psi$ are as discussed
in Example~\ref{pgxam2} and $\Gz$ is some subset of the global states,
then, as we observed above, $\Pg_2$ satisfies the specification ``$y$ is
never equal to 1''
for the initial condition $\INIT_1$ which can be characterized
by the formula $y=0$ but not for the initial condition $\INIT_2$
characterized by $x=0 \land y=0$.
Similarly, if $\Pg_3$ is the result of replacing the test $K(x=0)$ in
$\Pg_2$ by $\neg K(x=0)$, then $\Pg_3$ satisfies the liveness condition
``$y$ is eventually equal to 1'' for $\INIT_1$ but not for $\INIT_2$.
This shows that a standard
specification (in particular, one involving safety or liveness) may not
be monotonic with respect to maximal specification for a knowledge-based
program.
\exam

\xam\label{differ2}
Consider the standard program $\Pg_1$ again, but now consider a context
where there are two agents.  Intuitively, the
second agent never learns anything and plays no role.  Formally, this
is captured by taking the second agent's local state to always be
$\lambda$.  Thus, a global state now has the form $(\<a,b\>,\lambda)$.
We can again identify the global state with the local state of the
first agent (the one performing all the actions).  Thus, abusing
notation somewhat, we can consider the same set of contexts as in
Example~\ref{differ1}.  Now consider
the knowledge-based specification $K_2 (y = 0)$.  This is true
with respect to $\Gamma$ for the initial
condition $\INIT_1$ but not for $\INIT_2$.
This shows that even for a standard program,
a knowledge-based specification may not be monotonic
with respect to maximal satisfaction.
\exam

\xam\label{differ3}
In the muddy children problem discussed in \cite{HM1}, the
father of the children says ``Some [i.e., one or more] of you have mud
on your forehead.''  The father then repeatedly asks the children ``Do
you know that you have mud on your own forehead?''  Thus, the children
can be viewed as running a knowledge-based program according to which
a child answers ``Yes'' iff she knows that she has mud on her forehead.
The father's initial statement is taken
to restrict the possible initial global states to those where one or
more children have mud on their foreheads.
It is well known that, under this initial condition, the knowledge-based
program satisfies the liveness property ``all the children
with mud on their foreheads eventually know it''.  On the other hand, if
the father instead gives the children more initial information, by saying
``Child
1 has mud on his forehead''  (thus restricting the set of initial global
states to those where child 1 has mud on his forehead), none of the
children that have mud on
their forehead besides child 1 will be able to figure out that they have
mud on their forehead.  Roughly speaking, this is because the
information available to the children from child 1's ``No'' answer in
the original version of the story is no longer available once the
father gives the extra information.  (See \cite[Example 7.25]{FHMV}.)
This problem is not an artifact of using knowledge-based programs or
specifications.  Rather, it is really the case in the original puzzle
that if the father had said ``Child 1 has mud on his forehead'' rather
than ``Some of you have mud on  your foreheads'', the children with mud
on their foreheads would never be able to figure out that they had mud
on their foreheads.  Sometimes extra knowledge can be harmful!%
\footnote{Another example of the phenomenon
that extra knowledge can be harmful can be found in \cite{MDH}.  This is
also a well-known phenomenon in the economics/game theory
literature \cite{Neyman}.}
\exam

As should be clear from the preceding discussion, there are two notions
of monotonicity, which happen to coincide (and hold) for standard
programs and specifications, but differ if we consider knowledge-based
programs or knowledge-based specifications.
For knowledge-based programs and specifications, the
first notion of monotonicity holds, while the second (monotonicity
with respect to maximal satisfaction) does not.
Monotonicity is certainly a desirable property---%
for a monotonic specification and program, once we prove that the
specification holds for the program for a given initial condition, then
we can immediately conclude that it holds for all stronger
specifications.  Without monotonicity, one may have to reprove the
property for all stronger initial conditions.  Maximal
satisfaction also certainly seems like a reasonable generalization from
the standard case.  Thus, we should consider to what extent it
is a problem that we lose monotonicity for maximal satisfaction when we
consider knowledge-based programs and specifications.

Of course,
whether something is problematic is, in great measure, in the eye of
the beholder.  Nevertheless, I would claim that, in the case of
maximal satisfaction, the only properties
that are lost when the initial condition is strengthened
are either unimportant
properties, or properties that, roughly speaking, {\em ought\/} to be
lost.  More precisely, they are properties that happen to be true of
a particular context, but are not intrinsic properties of the program.
The examples and the technical discussion below should help to make the
point clearer.
Thus, this lack of monotonicity should not be viewed as a defect of
knowledge-based programs and specifications.  Rather, it correctly
captures the subtleties of knowledge acquisition in certain circumstances.

\section{Some examples}\label{examples}
Consider again
the program $\Pg_2$.  It can be viewed as saying ``perform a sequence of
actions (continually increasing $y$) if you know that
$x=0$''. In the system
$\Rrep(\Pg_2,\gamma')$,
the initial condition guarantees that the agent does not know the value
of $x$, and thus nothing is done. The strengthening of the
initial condition to $x=0 \land y=0$ described by $\gamma''$ guarantees
that the agent does know that $x=0$, and thus actions are performed.
In this
case, we surely do not want a safety condition like ``$y$ is never equal
to 1'', which holds if the sequence of actions is not performed,
to be preserved when we strengthen the initial condition in this way.
Similarly, for the program
$\Pg_3$ defined in Example~\ref{differ1}, where the action is performed
if the agent does not know that $x=0$, we would not expect a liveness
property like ``$y$ is eventually equal to 1'' to be preserved.

Clearly, there are times when we would like a safety or a liveness
property to be preserved when we strengthen initial conditions.
But these safety or liveness properties are typically ones that
we want to hold of {\em all\/} systems consistent with the
knowledge-based program, not just the ones representing the
program in certain maximal contexts.
 The tests in a well-designed
knowledge-based program are often there precisely to ensure that
desired safety properties do hold in all systems consistent with
the program.
For example, there may be a test for
knowledge to ensure that an action is
performed only if it is known to be safe (i.e.,~it does
not violate the safety property).
It is often possible to prove that such safety properties hold
in all systems consistent with the knowledge-based program; thus, the
issue of needing to reprove the property if we strengthen the initial
conditions
does not arise.  (See \cite[pp.~259--270]{FHMV} for further discussion
of this issue.)

In the case of liveness properties, we often want to ensure that a given
action is eventually performed.  It is
typically the case that an action in a knowledge-based program
is performed when a given fact is known to be true.  Thus, the
problem reduces to ensuring that the knowledge is eventually obtained.
As a consequence, the knowledge-based approach often makes it
clearer what is required for the liveness property to hold.
One example of how safety properties can be ensured by appropriate
tests for knowledge and how liveness properties reduce to showing
that a certain piece of knowledge is eventually obtained is given
by the knowledge-based programs of \cite{HZ}.
I illustrate these points here using a simpler example.

Suppose we have a network of $n$ processes, connected via a
communication network.  The network is connected, but not
necessarily completely connected.  For simplicity, assume
each communication link is bidirectional.  We assume that all messages
arrive within one time unit.
Each process knows
which processes it is connected to; formally, this means that
the local state of each process includes a mapping associating
each outgoing link with the identity of the neighbor at the other end.
We also assume that each process records
in its local state the messages it has sent and received.
We want a program for
process 1 to broadcast a binary value to
all the processes in the network.  Formally, we assume that each process
$i$ has a local variable, say $x_i$, which is intended to store the
value.  The specification that the program must satisfy consists of
three properties.
For every run, and for all $i = 1, \ldots, n$, we require the following:
\begin{enumerate}
\item
$x_i$ changes value at most once,
\item $x_1$ never
changes value, and
\item eventually the value of $x_i$ is equal to that of $x_1$.
\end{enumerate}
Note that the first two properties are safety properties, and
the last is a liveness property.

A simple standard program
that satisfies this specification is
for process 1 to send $v$, the value of $x_1$, to all
its neighbors; then the first time
process $i$ ($i \ne 1$)
gets the value $v$, it sets $x_i$ to $v$ and sends $v$
to all its neighbors
except the one from which it received the message.  Process $i$
does nothing
if it later gets the value $v$ again.
This program is easily seen to satisfy the specification in the context
implicitly described above.
We remark that, in principle, we could modify the first
property to allow $x_1$ to change value a number
of times before finally ``stabilizing'' on a final value.  However,
allowing
this would only complicate the description of the property,
since we would have to modify the third property to guarantee that
the value of $x_i$ after stabilizing is equal to that of $x_1$.  We
return to this point below.

The behavior of each process can
easily be captured in terms of knowledge:  When a process knows the
value of $x_1$, it sends the value to all its neighbors except
those that it knows already know the value of $x_1$.
Let $K_i(x_1)$ be an abbreviation for ``process $i$ knows
the value of $x_1$''.  (Thus, $K_i(x_1)$
is an abbreviation for $K_i(x_1 = 0)
\lor K_i(x_1 = 1)$.)  Similarly, let $K_i K_j (x_1)$ be an abbreviation
for ``process $i$ knows that process $j$ knows the value of $x_1$.''
Then we have
the joint knowledge-based program $\DIFF = (\DIFF_1, \ldots, \DIFF_n)$,
where $\DIFF_i$, the program followed by process
$i$, is
$$\begin{array}{l}
{\bf do\ forever}\\
\ \ \ \ \ {\bf if}\ K_i(x_1) \\
\ \ \ \ \ {\bf then} \\
\ \ \ \ \ \ \ \ \ \ x_i := x_1;\\
\ \ \ \ \ \ \ \ \ \ \mbox{{\bf for} each neighbor $j$ of $i$}\\
\ \ \ \ \ \ \ \ \ \ {\bf do}\\
\ \ \ \ \ \ \ \ \ \ \ \ \ \ \
{\bf if} \ \neg K_i K_j (x_1) \ {\bf then} \mbox{ send
the value of $x_1$ to $j$ {\bf end}}\\
\ \ \ \ \ \ \ \ \ \ {\bf end}\\
\ \ \ \ \ {\bf end}\\
{\bf end}.
\end{array}$$
By considering this knowledge-based program, we abstract away from
the details of how
$i$ gains knowledge of the value of $x_1$.
If $i=1$,
then presumably the value
was known all along; otherwise
it was perhaps acquired through the receipt of a message.
Similarly, the fact that $i$ sends the value of $x_1$ to a neighbor $j$ only if
$i$ doesn't know that $j$ knows the value of $x_1$
handles two of the details of the standard
program: (1)  it guarantees that $i$ does
not send the value of $x_1$ to $j$ if $i$ received the value of $x_1$ from $j$,
and (2) it guarantees that
$i$ does not send the value of $x_1$ to its neighbors more than once.%
\footnote{This argument depends in part on our assumption that
process $i$ is keeping track of the messages it sends
and receives.
If $i$ forgets the fact that it received the value of $x_1$
from $j$ then
(if $i$ follows $\DIFF_i$), it would send
the value of $x_1$ back to $j$.  Similarly, if $i$ receives the value of $x_1$ a second
time and forgets that it has already sent it once to its
neighbors, then according to $\DIFF_i$,
it would send it again.
In addition, the assumption that
there are no process failures is crucial.}
Finally, observe that $\DIFF$ is correct
even if messages can be lost, as long as the system satisfies an
appropriate fairness assumption (if a message is sent infinitely often,
it will eventually be delivered).%
\footnote{Note that this fairness assumption can be captured by using an
appropriate set $\Psi$ (consisting only of runs where
the fairness condition is satisfied) in the context.}
In this case process $i$ would keep
sending the value of $x_1$ to $j$ until $i$ knows (perhaps by receiving an
acknowledgment from $j$) that $j$ knows the value of $x_1$.
The fact that $\DIFF$ is correct
``even if messages can be lost'' or ``no matter what the network
topology'' means that the program meets its
specification in a number of different contexts.

This knowledge-based program has another advantage: it suggests
ways to design more efficient standard programs.  For example,
process $i$ does not have to send the value of $x_1$ to all its neighbors
(except the one from which it received the value of $x_1$) if it has
some other way of knowing that a neighbor already knows the value of $x_1$.
This may happen if the value of $x_1$ has a header describing
to which processes it has already been sent.  It might also happen
if the receiving process has some knowledge of the network topology
(for example, there is no need to rebroadcast the value of $x_1$ if
communication is reliable and all processes are neighbors of process 1).

Returning to our main theme, notice that in every context
$\gamma$ consistent with our assumptions, in the system(s) representing
$\DIFF$ in $\gamma$, the three
properties described above are satisfied:
$x_i$ changes value at most once in any run, $x_1$ never
changes value, and eventually the value of $x_i$ is equal to that of
$x_1$.
Notice also the role of the test $K_i(x_1)$ in ensuring that
the safety properties hold.  As a result of the test, we know that $x_i$
is not updated until the value of $x_1$ is known; when it is updated,
it is set to $x_1$.  This guarantees that $x_1$ never changes value,
and that $x_i$ changes value at most once and, when it does, it is
set to $x_1$.  All that remains is to guarantee that $x_i$ is eventually
set to $x_1$.  What the knowledge-based program makes clear is that
this amounts to ensuring that all processes eventually know the value
of $x_1$.  It is easy to prove that this is indeed the case.

It is also easy to see that
there are other properties that do not
hold in all contexts.
For a simple example, suppose that $n=3$, so there are three processes
in the network.  Suppose that there is a link from process 1 to process
2, and a link from process 2 to process 3, and that these are the
only links in the network.  Moreover, suppose that the network
topology is common knowledge.
Given these simplifying assumptions, a process $i$'s initial state
consists of an encoding of the network topology, its name, and the value
of $x_i$.  Now consider two contexts: in context $\gamma_1$, there are 8
initial global states, in which $(x_1,x_2,x_3)$ take on all values in
$\{0,1\}^3$; in $\gamma_2$, there are 4 initial global states, in which
$(x_1,x_2,x_3)$ take on all values in $\{0,1\}^3$ such that $x_1 = x_3$.
Intuitively, in context $\gamma_2$, process 3 knows the value of $x_1$
(since it is the same as the value of $x_3$, which is part of process
3's initial state), while in $\gamma_1$, neither process 2 nor process 3
know the value of $x_1$.  Let $\R_1 = \Rrep(\DIFF,\gamma_1)$ and let
$\R_2 = \Rrep(\DIFF,\gamma_2)$.
It is not hard to see that $\R_1$ has eight runs, one
corresponding to each initial global state.
In each of these runs,
process 1 sends the value of $x_1$ to
process 2 in round 1; process 2 sets $x_2$ to this value in round 2 and
forwards the value to process 3;
in round 3, process 3 sets $x_3$ to $i$
(and sends no messages).  (Note that, formally, {\em round\/} $k$ takes
place between times $k-1$ and $k$.)  Similarly, $\R_2$, has four runs,
one corresponding
to each initial global state.  In these runs, process 3 initially knows
the value of $x_1$, although process 2 does not.  Moreover, process 2
knows this.
Thus, in the round of the runs in $\R_2$, both process 1 and process 3
send the value of $x_1$ to process 2.  But now,
process 2 does not send a message to process 3 in the second round.

As expected, we can observe that not all liveness properties
are preserved as we move from $\R_1$ to $\R_2$.
For example, the runs in $\R_1$ all satisfy the liveness property
``eventually process 2 sends a message to process 3''.  Clearly the
runs in $\R_2$ do not satisfy this liveness property.
This should be seen as a feature, not a bug!  There is no reason
to preserve the sending of unnecessary messages.
The extra knowledge obtained when the initial conditions are
strengthened may render sending the message unnecessary.

\section{Discussion}\label{conclusion}
When designing programs, we often start with a specification and try to
find an (easily-implementable) standard program that satisfies it.
The process of going from a specification to an implementation is often
a difficult one.
I would argue that quite often it is useful to express the
properties we desire using a knowledge-based specification, proceed
from there to construct a knowledge-based program, and then go from the
knowledge-based program to a standard program.
While this approach may not
always be helpful (indeed, if a badly designed knowledge-based program
is used, then it may actually be harmful), there is some evidence
showing that it can help.

The first examples of going from knowledge-based specifications to
(standard) programs can be found in
\cite{APP,DM,Kurki} (although the
formal model used in \cite{APP,Kurki} is somewhat different from
that described here).
The approach described here was used in \cite{HZ} to derive solutions to
the {\em sequence transmission problem} (the problem of transmitting a
sequence of bits reliably over a possibly faulty communication channel).
All the programs derived in \cite{HZ}
are (variants of) well-known programs that solved the problem.
While I would argue that the knowledge-based approach shows the
commonality in the approaches used to solve the problem, and allows for
easier and more uniform proofs of correctness, certainly this example by
itself is not convincing evidence of the power of the knowledge-based
approach.

Perhaps more convincing evidence is provided by the results
of \cite{DM,HMW,MT}, where this approach is used to derive
programs that are optimal (in terms of number of rounds required) for
Byzantine Agreement and Eventual Byzantine Agreement.  In this case, the
programs derived were new, and it seems that it would have been quite
difficult to derive them directly from the original specifications.

Knowledge-based specifications are more prevalent than it might at
first seem.
We are often interested in constructing programs that
not only satisfy some safety and liveness conditions, but also
use a minimal number of messages or rounds.  As we have already
observed, specifications of the form ``do not send unnecessary
messages'' are not standard specifications; the same is true
for a specification of the form ``halt as soon as possible''.
Such specifications can be viewed as knowledge-based specifications.
The results of \cite{DM,HMW,MT} can be viewed as showing how
knowledge-based specifications
arise in the construction of round-efficient programs.
The tests for knowledge in the knowledge-based programs
described in these papers explicitly embody the intuition that a process
decides as soon as it is safe to do so.

Similar sentiments about the importance of knowledge-based specifications
are expressed by Mazer \citeyear{Maz91} (although the analogy between
knowledge-based
programs and knowledge-based specifications is not made in that paper):
\begin{quote}
Epistemic [i.e., knowledge-based] specifications are surprisingly common:
a problem specification that asserts that a property or value is
private to some process {\em is\/}
an epistemic specification (e.g., ``each
database site knows whether it has committed the transaction'').  We are
also interested in epistemic properties to capture assertions on the
extent to which a process's local state accurately reflects aspects of
the system state, such as ``each database site knows whether the others
have committed the transaction''.
\end{quote}

For another example of the usefulness of knowledge-based specifications,
recall our earlier discussion of the specification of the program
for broadcasting a message through a network.  If we replace the
liveness requirements by the simple knowledge-based requirement
``eventually process $i$ knows the value of $x_1$'', we can drop
the first property (that $x_i$ changes value at most once)
altogether.  Indeed,
we do not have to mention $x_i$, $i \ne 1$, at all.  The knowledge-based
specification thus seems to capture our intuitive requirements for
the program more directly and elegantly
than the standard specification given.

A standard specification can be viewed as a special case of a
knowledge-based specification, one in which the set of systems satisfying
it is closed under unions and subsets.
It is because of these closure properties that we have the property if a
standard program satisfies a standard specification $\sigma$ in a context
$\gamma$, then it satisfies it in any restriction of $\gamma$.
Clearly, this is not a property that holds of standard programs once we
allow knowledge-based specifications.  Nevertheless, as the examples
above
suggest, there is something to be gained---and little to be lost---by
allowing the greater generality of knowledge-based specifications.
In particular, although we do lose monotonicity, there are other ways of
ensuring that safety and liveness properties do hold in the systems of
interest.

By forcing us to think in terms of systems, rather than of individual
runs, both knowledge-based programs and knowledge-based specifications
can be viewed as requiring more ``global'' thinking than their
standard counterparts. The hope is that thinking at this level
of abstraction
makes the design and specification of programs easier to carry out.

We still need more experience using this framework
before we can decide whether this hope will be borne out
and whether the knowledge-based approach as described here is really
useful.
Sanders has other criticisms of the use of knowledge-based programs
that I have not addressed here.  Very roughly, she provides
pragmatic arguments that suggest that we use predicates that
have some of the properties of knowledge (for example $K \phi \rimp
\phi$), but not necessarily all of them.  This theme is further pursued
in \cite{EMM98}.  While I believe that using predicates that satisfy
some of the properties of knowledge will not prove to be as useful as
sticking to the original notion of knowledge, we clearly need more
examples to better understand the issues.

 Besides more examples,
as pointed out by Sanders \citeyear{Sanders}, it would also be useful to
have techniques
for reasoning about knowledge-based programs without having to construct
the set of runs generated by the program.
In \cite{FHMV}, a simple knowledge-based programming language is
proposed.  Perhaps standard
techniques for proving program correctness can be applied to it (or some
variant of it).   A first step along these lines was taken by Sanders
\citeyear{Sanders}, who extended {\sc UNITY} \cite{CM88} in such a way as to
allow the definition of knowledge predicates (although it appears that
the resulting knowledge-based programs are somewhat less general
than those described here), and then used proof techniques developed
for {\sc UNITY} to prove the correctness of another
knowledge-based protocol for the sequence transmission problem.
(We remark that techniques for reasoning about knowledge obtained
in CSP programs, but not for knowledge-based programs, were
given in \cite{KT}.)
Once we have
a number of examples and better techniques in hand, we shall need
to carry out a careful evaluation of the knowledge-based approach,
and a comparison of it and other approaches.
I believe that once the evidence is in, it
will show that there are indeed significant advantages
that can be gained by thinking at the knowledge level.
\bigskip

\noindent{\bf Acknowledgments:}  I would like to thank Ron Fagin,
Yoram Moses, Beverly Sanders, and particularly
Vassos Hadzilacos, Murray Mazer, Moshe Vardi, and Lenore
Zuck for their helpful comments on earlier drafts of the paper.  Moshe
gets the credit for the observation that knowledge-based protocols
do satisfy monotonicity.  Finally, I would
like to thank Karen Seidel for asking a question at PODC '91 that
inspired this paper.

\bibliographystyle{plain}
\bibliography{z,joe}
\end{document}

%% file: defn.tex
\newtheorem{THEOREM}{Theorem}[section]
\newenvironment{theorem}{\begin{THEOREM} \hspace{-.85em} {\bf :} }%
                        {\end{THEOREM}}
\newtheorem{LEMMA}[THEOREM]{Lemma}
\newenvironment{lemma}{\begin{LEMMA} \hspace{-.85em} {\bf :} }%
                      {\end{LEMMA}}
\newtheorem{COROLLARY}[THEOREM]{Corollary}
\newenvironment{corollary}{\begin{COROLLARY} \hspace{-.85em} {\bf :} }%
                          {\end{COROLLARY}}
\newtheorem{PROPOSITION}[THEOREM]{Proposition}
\newenvironment{proposition}{\begin{PROPOSITION} \hspace{-.85em} {\bf :} }%
                            {\end{PROPOSITION}}
\newtheorem{DEFINITION}[THEOREM]{Definition}
\newenvironment{definition}{\begin{DEFINITION} \hspace{-.85em} {\bf :} \rm}%
                            {\end{DEFINITION}}
\newtheorem{CLAIM}[THEOREM]{Claim}
\newenvironment{claim}{\begin{CLAIM} \hspace{-.85em} {\bf :} \rm}%
                            {\end{CLAIM}}
\newtheorem{EXAMPLE}[THEOREM]{Example}
\newenvironment{example}{\begin{EXAMPLE} \hspace{-.85em} {\bf :} \rm}%
                            {\end{EXAMPLE}}
\newtheorem{REMARK}[THEOREM]{Remark}
\newenvironment{remark}{\begin{REMARK} \hspace{-.85em} {\bf :} \rm}%
                            {\end{REMARK}}

\newcommand{\thm}{\begin{theorem}}
\newcommand{\lem}{\begin{lemma}}
\newcommand{\pro}{\begin{proposition}}
\newcommand{\dfn}{\begin{definition}}
\newcommand{\rem}{\begin{remark}}
\newcommand{\xam}{\begin{example}}
\newcommand{\cor}{\begin{corollary}}

\newcommand{\ethm}{\end{theorem}}
\newcommand{\elem}{\end{lemma}}
\newcommand{\epro}{\end{proposition}}
\newcommand{\edfn}{\bbox\end{definition}}
\newcommand{\erem}{\bbox\end{remark}}
\newcommand{\exam}{\bbox\end{example}}
\newcommand{\ecor}{\end{corollary}}

\newcommand{\beqn}{\begin{equation}}
\newcommand{\eeqn}{\end{equation}}

\newcommand{\bbox}{\vrule height7pt width4pt depth1pt}

\newcommand{\clm}{\begin{claim}}
\newcommand{\eclm}{\end{claim}}

\newcommand{\rimp}{\Rightarrow}

\renewcommand{\phi}{\varphi}

\newcommand{\G}{{\cal G}}

\newcommand{\R}{{\cal R}}

\renewcommand{\>}{\rangle}

\newcommand{\ol}{\setlength{\itemsep}{0pt}\begin{enumerate}}
\newcommand{\eol}{\end{enumerate}\setlength{\itemsep}{-\parsep}}
\newcommand{\ul}{\setlength{\itemsep}{0pt}\begin{itemize}}
\newcommand{\dl}{\setlength{\itemsep}{0pt}\begin{description}}
\newcommand{\edl}{\end{description}\setlength{\itemsep}{-\parsep}}
\newcommand{\eul}{\end{itemize}\setlength{\itemsep}{-\parsep}}

\newcommand{\commentout}[1]{}

\newcommand{\bi}{\begin{itemize}}
\newcommand{\ei}{\end{itemize}}
\newcommand{\be}{\begin{enumerate}}
\newcommand{\ee}{\end{enumerate}}

\newcommand{\Gz}{\G_0}